\documentclass[10pt,twocolumn,showpacs,amsmath,amssymb,floatfix,superscriptaddress]{revtex4-2}
\usepackage{graphicx}
\usepackage{float}
\usepackage{dcolumn}
\usepackage{bm}
\usepackage{color}
\usepackage{txfonts}
\usepackage{microtype}
\usepackage{hyperref}
\usepackage{mathrsfs}
\usepackage{easyReview}
\usepackage{amsmath}
\usepackage{extarrows}
\usepackage{mathtools}
\usepackage{xcolor}
\usepackage{slashed}

\setreviewson
\begin{document}

\title{Vortex $\gamma$ photon generation via spin-to-orbital angular momentum transfer in nonlinear Compton scattering}

\author{Mamutjan Ababekri}
\affiliation{Ministry of Education Key Laboratory for Nonequilibrium Synthesis and Modulation of Condensed Matter, Shaanxi Province Key Laboratory of Quantum Information and Quantum Optoelectronic Devices, School of Physics, Xi'an Jiaotong University, Xi'an 710049, China}
\author{Ren-Tong Guo}
\affiliation{Ministry of Education Key Laboratory for Nonequilibrium Synthesis and Modulation of Condensed Matter, Shaanxi Province Key Laboratory of Quantum Information and Quantum Optoelectronic Devices, School of Physics, Xi'an Jiaotong University, Xi'an 710049, China}
\author{Feng Wan}\email{wanfeng@xjtu.edu.cn}
\affiliation{Ministry of Education Key Laboratory for Nonequilibrium Synthesis and Modulation of Condensed Matter, Shaanxi Province Key Laboratory of Quantum Information and Quantum Optoelectronic Devices, School of Physics, Xi'an Jiaotong University, Xi'an 710049, China}
\author{B. Qiao}
\affiliation{Center for Applied Physics and Technology, HEDPS and SKLNPT, School of Physics, Peking University, Beijing 100871, China}
\author{Zhongpeng Li}
\affiliation{Ministry of Education Key Laboratory for Nonequilibrium Synthesis and Modulation of Condensed Matter, Shaanxi Province Key Laboratory of Quantum Information and Quantum Optoelectronic Devices, School of Physics, Xi'an Jiaotong University, Xi'an 710049, China}
\author{Chong Lv}
\affiliation{Department of Nuclear Physics, China Institute of Atomic Energy, P. O. Box 275(7), Beijing 102413, China}
\author{Bo Zhang}
\affiliation{Key laboratory of plasma physics, Research center of laser fusion, China academy of engineering physics, 621900, Mianshan Road 64\#, Mianyang, Sichuan, China}
\author{Weimin Zhou}
\affiliation{Key laboratory of plasma physics, Research center of laser fusion, China academy of engineering physics, 621900, Mianshan Road 64\#, Mianyang, Sichuan, China}
\author{Yuqiu Gu}
\affiliation{Key laboratory of plasma physics, Research center of laser fusion, China academy of engineering physics, 621900, Mianshan Road 64\#, Mianyang, Sichuan, China}
\author{Jian-Xing Li}\email{jianxing@xjtu.edu.cn}
\affiliation{Ministry of Education Key Laboratory for Nonequilibrium Synthesis and Modulation of Condensed Matter, Shaanxi Province Key Laboratory of Quantum Information and Quantum Optoelectronic Devices, School of Physics, Xi'an Jiaotong University, Xi'an 710049, China}
\date{\today}

\begin{abstract}
Vortex $\gamma$ photons with intrinsic orbital angular momenta (OAM) possess a wealth of applications in various fields, e.g.---strong-laser physics, nuclear physics, particle physics and astrophysics---yet their generation remains unsettled.  In this work, we investigate the generation of vortex $\gamma$ photons via nonlinear Compton scattering of ultrarelativistic electrons in a circularly polarized laser pulse.  
We develop a quantum electrodynamics scattering theory that explicitly addresses the multiphoton absorption and the angular momentum transfer mechanism.  
In pulsed laser fields, we unveil the vortex phase structure of the scattering matrix element, discuss how the vortex phase could be transferred to the radiated photon, and derive the radiation rate of the vortex $\gamma$ photon.  We numerically examine the energy spectra and beam characteristics of the radiation, while also investigating the influence of finite laser pulses on the angular momentum and energy distribution of the emitted vortex $\gamma$ photons.   
%Moreover, the nonlinear Breit-Wheeler scattering of the vortex $\gamma$ photon in a strong laser can reveal the phase structures of the vortex $\gamma$ photons.  
%Our findings emphasize the special role played by the intense laser regarding both generation and call for further investigations.  
\end{abstract}

\maketitle

\section{introduction}

Vortex photons are described by complex wave functions with helical phases, and they carry intrinsic orbital angular momentum (OAM) along the propagation direction \cite{allen1992orbital,knyazev2018beams}.  
Light beams with OAM have significant applications in various fields including optical manipulation \cite{he1995direct,garces2003observation}, quantum optics \cite{mair2001entanglement,leach2009violation}, imaging \cite{swartzlander2001peering,swartzlander2008astronomical}, etc.  
Due to their new degree of freedom, high energy vortex photons offer an alternative approach for spin physics and promise to shed new light on nuclear and particle physics \cite{ivanov2020doing,ivanov2022promises}.  
Experimental opportunities introduced by vortex photons and their collisions with ion beams are studied regarding existing and future physics programs \cite{durante2019all,krasny2015gamma,budker2020atomic,budker2021arxiv}.  Vortex photons have also been shown to reveal various quantum electrodynamics (QED) effects in \cite{aboushelbaya2019orbital,sherwin2017compton,lei2021generation,bu2021twisted}.   
Currently, vortex photons from visible to x-ray (eV $\sim$ keV) regimes have been generated utilizing optical mode conversion techniques, high harmonic generation or coherent radiation in helical undulators and laser facilities \cite{shen2019optical,peele2002observation,terhalle2011generation,gariepy2014creating,hemsing2013coherent}.  However, the generation of vortex $\gamma$ photons ($\gtrsim$ MeV) still remains a challenge, with most studies considering Compton scattering to attain high energy and OAM \cite{jentschura2011generation,petrillo2016compton,taira2017gamma,chen2019generation,chen2019generation,bogdanov2019semiclassical}.  

Ultraintense and ultrashort laser pulses \cite{danson2019petawatt,yoon2019achieving,yoon2021realization} have been demonstrated to generate high energy $\gamma$ photons via nonlinear Compton scattering (NCS) \cite{chen2013mev,sarri2014ultrahigh,cole2018experimental,poder2018experimental}.   
The spin angular momentum (SAM) properties of emitted $\gamma$ photons have been revealed in theoretical investigations from moderate to ultraintense regimes \cite{yanfei2020polarized,tang2020highly,wang2022brilliant,Sun:2022cat}.  
The generated $\gamma$ photon beam is shown to carry OAM as the collective effect of the mechanical OAM of the individual photon defined by $\bm{L}=\bm{r}\times\bm{p}$ \cite{liu2016generation,Chen:2018tkb}.  
Although these studies have inspired rich mechanisms for the generation of vortex beams with collective OAM in laser-plasma interactions \cite{gong2018brilliant,liu2020vortex,wang2020generation,hu2021attosecond,zhang2021efficient,bake2022bright}, the intrinsic OAM carried by these photons is still missing from the picture.    
On the other hand, investigations based on the spiral motion of charged particles in laser or undulator fields have identified that photons in harmonic radiation carry intrinsic OAM \cite{sasaki2008proposal,bahrdt2013first,katoh2017angular}.  
Recently, the generation of vortex particles during a scattering process was found to be the result of a generalized measurement over the final particle in the quantum picture \cite{karlovets2022generation}. Thus, radiation of vortex $\gamma$ photons should be investigated in the full QED treatment to reflect the quantum evolution of the electron, which is missing in classical trajectory-based studies \cite{taira2017gamma,katoh2017angular,bogdanov2018probability,epp2019angular,bogdanov2019semiclassical}. 
Moreover, strong lasers are achieved by temporal compression, and the resulting finite pulse shape alters the harmonic content of the NCS spectra affecting angular momentum properties of radiated $\gamma$ photons \cite{mackenroth2011nonlinear,tang2020highly}.  

During the NCS process, each radiation event turns multiple low energy laser photons into a single high energy final photon such that the angular momentum transfer relation cannot be addressed if one only considers SAM.  
Unfortunately, the standard QED treatment of the radiation mostly investigates SAM as the only internal degree of freedom affecting the process \cite{gonoskov2022charged,fedotov2022advances}.  
Thus, the theory fails to grasp the angular momentum conservation property of radiation events when the system respects rotational invariance, which could be the case in circularly polarized (CP) laser fields.    
Moreover, as the radiation under rotational symmetry could be realized in upcoming intense laser-based experiments \cite{abramowicz2021conceptual,salgado2021single,meuren2020seminal}, the vortex nature of strong-field QED processes calls for detailed investigations.  

\begin{figure}[!t]
	\setlength{\abovecaptionskip}{-0.0cm}
	\includegraphics[width=1.0\linewidth]{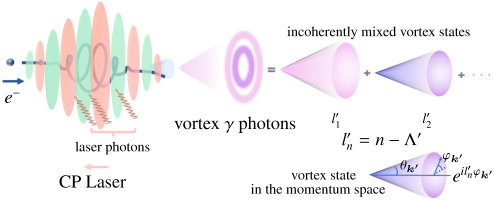}
	\begin{picture}(300,0)	
	\end{picture}
	\caption{ Generation of vortex $\gamma$ photons with intrinsic OAM in NCS of the ultrarelativistic electron in the CP laser pulse. The OAM of the vortex $\gamma$ photon is  $l^\prime_n=n-\Lambda^\prime$, where $n$ labels the corresponding harmonics and $\Lambda^\prime$ is the $\gamma$ photon spin. The vortex state is represented by a cone in momentum space with the cone angle $\theta_{\bm{k}^{\prime}}$ and the azimuth angle $\varphi_{\bm{k}^{\prime}}$ of the momentum vector $\bm{k}^\prime$.  }
	\label{fig_illustration}
\end{figure}

In this study, we investigate the generation of vortex $\gamma$ photons with intrinsic OAM during the NCS process of ultrarelativistic electrons and an intense CP laser pulse. 
As presented in Fig.~\ref{fig_illustration}, a plane wave electron propagates along the $z$ direction, collides with a counterpropagating plane wave CP laser pulse, and emits vortex photons. 
We develop a vortex scattering theory within the framework of the Furry picture of strong-field QED to describe the radiation process. 
We derive analytical expressions for the wave function of the radiated photon and the corresponding radiation probability rate. Moreover, we conduct numerical investigations to explore the energy spectra and beam properties of the emitted vortex $\gamma$ photon.

Our paper is organized as follows: In Sec. II, we present our theoretical framework by unveiling the vortex phase structure of the $S$-matrix element, obtaining the radiated photon wave function, and deriving the differential probability for the radiation of the vortex $\gamma$ photon. In Sec. III, we present numerical results for the differential radiation rate and the beam property of the vortex $\gamma$ photon. We conclude with final remarks in Sec. IV.
Throughout, natural units are used ($\hbar=c=1$), the fine-structure constant is $\alpha=\frac{e^2}{4\pi}\approx\frac{1}{137}$ with the electron charge $e=-\vert e\vert$, and the electron mass is denoted as $m_e$.

\section{Theoretical Framework}
Vortex states represent non-plane wave solutions to respective wave equations.
When expressed in cylindrical coordinates $r$, $\varphi$ and $z$, a vortex scalar particle can be described using the following wave function \cite{jentschura2011compton}:
\begin{equation}
	\psi_{\varkappa m k_z}(r,\varphi_r,z,t)=\frac{e^{-i(\omega t - k_z z)}}{\sqrt{2 \omega}}\psi_{\varkappa m}(r,\varphi_r)\,,
	\label{vortex_state}
\end{equation}
with the wave form function
\begin{equation}
	\psi_{\varkappa m}(r,\varphi_r)=\sqrt{\frac{\varkappa}{2 \pi}} e^{-i m \varphi} J_{m}(\varkappa r)\,,
\end{equation}
where $J_m$ is the Bessel function of the first kind. Expression \eqref{vortex_state} represents the Bessel vortex state, which is considered as the eigenfunction of the $z$ projection of the OAM operator $\hat{L}_z=-i\partial/\partial \varphi_r$ with the corresponding eigenvalue $m$.  Therefore, a vortex scalar particle can be characterized by its energy $\omega$, longitudinal momentum $k_z$, transverse momentum $\varkappa$, and the projection of OAM onto the $z$ axis $m$.
The vortex state could be written in terms of plane wave states:
\begin{equation}
	\psi_{\varkappa m k_z}(r,\varphi_r,z,t)= \int \frac{d^2 \bm{k}_\perp}{(2\pi)^2}\,a_{\varkappa\,m}(\bm{k}_\perp)\,\frac{e^{-i(\omega t - \bm{k\cdot x})}}{\sqrt{2 \omega}}\,,
\label{vortex_pw}
\end{equation}
where the Fourier amplitude $a_{\varkappa\,m}(\bm{k}_\perp)$ encodes a cone structure with radius $k_\perp=\vert \bm{k}_\perp \vert =\varkappa$ in the momentum space with a helical phase factor $e^{i\,m\varphi_{\bm{k}}}$:
\begin{equation}
	a_{\varkappa\,m}(\bm{k}_\perp)=(-i)^m\,e^{i\,m\varphi_{\bm{k}}}\,\sqrt{\frac{2\pi}{\varkappa}}\delta(k_\perp-\varkappa)\,.
\label{vortex_amp}
\end{equation}  
Similarly, the wave function of the vortex photon or electron can also be expressed as a superposition of corresponding plane wave states using the Fourier amplitude $a_{\varkappa\,m}(\bm{k}_\perp)$, as is done in Eq.~\eqref{vortex_pw}.  Therefore, the scattering matrix element $S_{fi}^\text{vortex}$ involving such particles is naturally connected to the conventional plane wave scattering matrix element $S_{fi}^\text{plane}$ through the Fourier amplitudes in Eq.~\eqref{vortex_amp}.  
In the subsequent discussion, we will begin with the NCS $S$-matrix element for plane wave electrons and photons. Then, utilizing the result, we will proceed to calculate the scattering process involving vortex particles.  

\subsection{$S$-matrix element for the NCS in the pulsed laser}
The CP pulsed laser is modeled by the plane wave background field:
\begin{eqnarray}
\begin{aligned}
A(\phi)=a\,g(\phi)
\begin{pmatrix} 0 \\ \cos\phi \\ \sin\phi \\ 0 \end{pmatrix}=\frac{a}{\sqrt{2}}g(\phi)(e^{-i\phi}\epsilon_{+}+e^{+i\phi}\epsilon_{-})\,,
\end{aligned}\label{laser_field}
\end{eqnarray}
where $\phi=k\cdot x$ represents the laser phase with the wave 4-vector $k=(\omega,0,0,-\omega)$ associated with the laser photon, $g(\phi)$ is the pulse shape function, and $a=\xi\frac{m_e}{e}$ is defined by the dimensionless parameter $\xi$, which is related to the laser intensity through $\xi\approx\sqrt{7.3\times10^{-19}I[\text{W/cm}^2]\lambda_0^2[\mu\text{m}]}$. 
 The individual laser photons associated with the classical CP plane wave background field in Eq.~(\ref{laser_field}) carry SAM $s_{z}=+1$ along the $z$ axis. This can be seen by examining the canonical SAM density $\bm{E}\times\bm{A}$ using classical fields corresponding to the laser field \cite{Leader:2013jra}.
The representation of the laser field with polarization vectors $\epsilon_{\pm}^{\mu}=\frac{1}{\sqrt{2}}(0,1,\pm i,0)^{T}$ is introduced for later convenience when we define the dynamical integral $\mathscr{A}_{\pm}(s)$ in Eq.~\eqref{dynamic_integral}.  

We calculate the NCS process of an electron interacting with a pulsed laser in the Furry picture \cite{fedotov2022advances}, in which we treat the interaction between the laser field and the electron exactly, while considering the radiation perturbatively.  To describe the electron, we use the Volkov solution of the Dirac equation in the laser field \cite{Berestetskii1982quantum}, which, after introducing some notations such as $\Omega_p=\frac{\omega m_e}{2k\cdot p}\xi$, $A(\phi)=a\,\tilde{A}(\phi)$ and $k^\mu=\omega n^\mu$, can be written as
\begin{equation}
\psi_{p,\lambda}(x)=(1+\Omega_p\slashed{n}\slashed{\tilde{A}}(\phi))u_{p,\lambda}e^{-iS_p(\phi)}. 
\label{volkov_state}
\end{equation}
The classical action of the electron in the laser field in Eq.~\eqref{volkov_state} is
\begin{equation}
\begin{aligned}
S_{p}(\phi)&=p\cdot x+\frac{e}{k\cdot p}\int^\phi~d\phi^\prime (p\cdot A(\phi^\prime)-\frac{e}{2}A^2(\phi^\prime))\,, \\
&=p\cdot x+\alpha_{p_{\perp}}g(\phi)\sin(\varphi_{\bm{p}}-\phi)+\beta_{p}\int^\phi d\phi^\prime\,g^2(\phi^\prime)\,,
\end{aligned}
\label{classical_action}
\end{equation}
where $\varphi_{\bm{p}}$ is the azimuth angle of the momentum vector $\bm{p}$, we have introduced the notations $\alpha_{p_{\perp}}=\frac{p_{\perp} m_e}{k\cdot p}\xi$ and $\beta_p=\frac{m_e^{2}}{2k\cdot p}\xi^{2}$, and in the derivation of the second line, we have assumed the slowly varying envelope approximation $\frac{\partial}{\partial \phi} g(\phi) \ll g(\phi)$ allowing us to approximate $\int d\phi g(\phi) \cos(\phi) \approx g(\phi) \sin(\phi)$ \cite{seipt2016analytical}.   
Under the condition of switching off the laser field ($\xi=0$), the Volkov state given by Eq.~(\ref{volkov_state}) simplifies to the familiar wave function of a Dirac spinor $\psi_{p,\lambda}(x)=u_{p,\lambda}e^{-ip\cdot x}$, which describes a free electron.

The leading order $S$-matrix element for NCS with plane wave particles can be written as
\begin{equation}
\begin{aligned}
S_{fi}=&-ie\int~d^4x\,\bar{\psi}_{p^{\prime},\lambda^\prime}(x)\slashed{A}^{\prime \ast}_{k^\prime,\Lambda^\prime}(x)\psi_{p,\lambda}(x)\,,\\
=&-ie\int\,d^4x\,\bar{u}_{p^\prime,\lambda^\prime}[\slashed{\epsilon}^{\prime\ast}_{k^\prime,\Lambda^\prime}+\Omega_p\Omega_{p^\prime}g^2(\phi)(2n\cdot \epsilon^{\prime\ast}_{k^\prime,\Lambda^\prime})\slashed{n}\,\\
&+\Omega_{p^\prime}\slashed{\tilde{A}}(\phi)\slashed{n}\slashed{\epsilon}^{\prime\ast}_{k^\prime,\Lambda^\prime}+\Omega_{p}\slashed{\epsilon}^{\prime\ast}_{k^\prime,\Lambda^\prime}\slashed{n}\slashed{\tilde{A}}(\phi)]u_{p,\lambda}\,\\
&\times e^{-i(p-p^\prime-k^\prime)x}\,e^{-i\alpha_{p^\prime_{\perp}}g(\phi)\sin(\phi-\varphi_{\bm{p}^\prime})}\,e^{-i\Delta\beta\int^\phi d\phi^\prime g^2(\phi^\prime)}\,,
\end{aligned}
\label{S_fi_NCS}
\end{equation}
where $A^{\prime\mu}(x)={\epsilon}^{\prime\mu}_{k^\prime,\Lambda^\prime}~e^{-ik^{\prime}\cdot~x}$ is the radiated photon wave function, and $\Delta\beta=(\frac{1}{2k\cdot p}-\frac{1}{2k\cdot p^\prime})m_e^{2}\xi^{2}$. The familiar notations $\bar{\psi}=\psi^\dagger\gamma^0$ and $\slashed{a}=a_\mu\gamma^\mu$ are employed. The momentum $4$-vectors for the incoming electron, outgoing electron and emitted photon are denoted as $p=(\varepsilon,0,0,p_z)$, $p^{\prime}=(\varepsilon^{\prime},\bm{p}^\prime)$, and  $k^{\prime}=(\omega^\prime,\bm{k}^\prime)$, respectively.  The spin indices for the initial (final) electron and the emitted photon are represented by $\lambda$ ($\lambda^\prime$) and $\Lambda^\prime$ respectively. 

To obtain the energy-momentum conserving delta function, we perform a Fourier transform by rewriting the $S$-matrix element as $S_{fi}\propto\int d^4x \frac{ds}{2\pi} \ldots e^{-is\phi}e^{-i(p-p^\prime-k^\prime)x}$, which leads to the following dynamic integrals:
\begin{eqnarray}
	\begin{aligned}
		\mathscr{A}_{0}(s)&=\int d\phi\, e^{is\phi-i\alpha_{p_{\perp}^{\prime}}g(\phi)\sin(\phi-\varphi_{\bm{p}^{\prime}})-i\Delta\beta \int ^{\phi}d \phi^{\prime}g^{2}(\phi^{\prime})}\,,\\
		\mathscr{A}_{2}(s)&=\int d\phi\,g^{2}(\phi) e^{is\phi-i\alpha_{p_{\perp}^{\prime}}g(\phi)\sin(\phi-\varphi_{\bm{p}^{\prime}})-i\Delta\beta \int ^{\phi}d \phi^{\prime}g^{2}(\phi^{\prime})}\,,\\
		\mathscr{A}_{\pm}(s)&=\int d\phi\,g(\phi) e^{i(s\mp 1)\phi-i\alpha_{p_{\perp}^{\prime}}g(\phi)\sin(\phi-\varphi_{\bm{p}^{\prime}})-i\Delta\beta \int ^{\phi}d \phi^{\prime}g^{2}(\phi^{\prime})}\,.
	\end{aligned}\label{dynamic_integral}
\end{eqnarray}
By defining the charge currents as 
\begin{eqnarray}
	\begin{aligned}	
		\mathcal{J}_{0}&=\overline{u}_{p^{\prime},\lambda^{\prime}}\slashed{\epsilon}_{k^{\prime},\Lambda^{\prime}}^{\ast}u_{p,\lambda}\,,\quad
		\mathcal{J}_{2}=\Omega_{p}\Omega_{p^{\prime}}(2n\cdot\epsilon_{k^{\prime},\Lambda^{\prime}}^{\ast})\overline{u}_{p^{\prime},\lambda^{\prime}}\slashed{n}u_{p,\lambda}\,,\nonumber\\
		\mathcal{J}_{\pm}&=\frac{1}{\sqrt{2}}\overline{u}_{p^{\prime},\lambda^{\prime}}(\Omega_{p^{\prime}}\slashed{\epsilon}_{\pm}\slashed{n}\slashed{\epsilon}_{k^{\prime},\Lambda^{\prime}}^{\ast}+\Omega_{p}\slashed{\epsilon}_{k^{\prime},\Lambda^{\prime}}^{\ast}\slashed{n}\slashed{\epsilon}_{\pm})u_{p,\lambda}\,,\nonumber\\
	\end{aligned}
\end{eqnarray}
the $S$-matrix element can be written as
\begin{equation}
\begin{aligned}
S_{fi}&=-ie\int ds (2\pi)^3\delta^{(4)}(p+sk-p^\prime-k^\prime)\sum_{\sigma=0,2,\pm}\mathscr{A}_{\sigma}(s)\,\mathcal{J}_{\sigma}\\
&=-ie\frac{(2\pi)^{3}}{\omega} \delta^{(3)}(\bm{p}+s_0\bm{k}-\bm{p}^{\prime}-\bm{k}^{\prime})\sum_{\sigma=0,2,\pm}\mathscr{A}_{\sigma}(s_{0})\,\mathcal{J}_{\sigma} \,.	\end{aligned}
\label{S_fi_NCS}
\end{equation}
Here, the laser photon absorption number $s_0=\frac{ \varepsilon^{\prime} + \omega^{\prime}-\varepsilon}{\omega}$ takes continuous values, which arise from quantifying the number of absorbed laser photons with the central laser photon energy $\omega$. It should be noted that laser photons have a broad spectrum due to the finite pulse duration.

\subsection{Vortex phase structure}
Let us now exploit the azimuth angle dependency of the $S$-matrix element, as it is crucial for the discussion related to OAM.   
First, we need to extract the azimuth angle contributions due to the multiple laser photon absorption which is hidden in the dynamic integrals in Eq.~(\ref{dynamic_integral}). This is achieved by employing the Jacobi-Anger relation $e^{i\alpha g(\phi)\sin(\phi + \varphi_{\bm{p}})}=\sum_n J_n(\alpha g(\phi)) {e^{i n ({\phi} + \varphi_{\bm{p}})}}$ and rearranging the resulted summations, which leads to 
\begin{eqnarray}
	\begin{aligned}
		\mathscr{A}_{0}(s)&=\sum_{n}\int d\phi\,e^{i(s-n)\phi-i\beta \int ^{\phi}d \phi^{\prime}g^{2}(\phi^{\prime})}J_{n}(\alpha_{p_{\perp}^{\prime}}g(\phi))e^{in\varphi_{\bm{p}^{\prime}}}\nonumber\\
		&\equiv\sum_{n} \mathscr{C}_{0}^{(n)}(s)\,e^{in\varphi_{\bm{p}^{\prime}}}\,,\nonumber\\
		\mathscr{A}_{2}(s)&=\sum_{n}\int d\phi\,e^{i(s-n)\phi-i\beta \int ^{\phi}d \phi^{\prime}g^{2}(\phi^{\prime})}g^{2}(\phi)J_{n}(\alpha_{p_{\perp}^{\prime}}g(\phi))e^{in\varphi_{\bm{p}^{\prime}}}\nonumber\\
		&\equiv\sum_{n} \mathscr{C}_{2}^{(n)}(s)\,e^{in\varphi_{\bm{p}^{\prime}}}\,,\nonumber\\
		\mathscr{A}_{\pm}(s)&=\sum_{n}\int d\phi\,e^{i(s-n)\phi-i\beta \int ^{\phi}d \phi^{\prime}g^{2}(\phi^{\prime})}g(\phi)J_{n\mp1}(\alpha_{p_{\perp}^{\prime}}g(\phi))e^{i(n\mp1)\varphi_{\bm{p}^{\prime}}}\nonumber\\
		&\equiv\sum_{n}\mathscr{C}_{\pm}^{(n)}(s)\,e^{i(n\mp1)\varphi_{\bm{p}^{\prime}}}\,.
	\end{aligned}\label{dyn_int_har}
\end{eqnarray}
At this point, the $S$-matrix element is expanded into harmonics as follows:
\begin{eqnarray}
	\begin{aligned}
		S_{fi}&=-ie\frac{(2\pi)^{3}}{\omega}\delta^{(3)}(\bm{p}+s_{0}\bm{k}-\bm{p}^{\prime}-\bm{k}^{\prime})\sum_{n}\\
		&\times\{e^{in\varphi_{\bm{p}^{\prime}}}[\mathscr{C}_{0}^{(n)}(s_{0})\mathcal{J}_{0}+\mathscr{C}_{2}^{(n)}(s_{0})\mathcal{J}_{2}]+e^{i(n-1)\varphi_{\bm{p}^{\prime}}}\mathscr{C}_{+}^{(n)}(s_{0})\mathcal{J}_{+}\\
		&+e^{i(n+1)\varphi_{\bm{p}^{\prime}}}\mathscr{C}_{-}^{(n)}(s_{0})\mathcal{J}_{-}\} \,.
	\end{aligned}\label{harmonic_S}
\end{eqnarray}
It is important to note that the discrete number $n$ labeling the harmonics and the continuous number $s_0$ denoting the photon absorption number reflect the laser photon contributions during the radiation process. The discreteness of $n$ is associated with the fact that each laser photon, even with a broad energy distribution due to the finite pulse shape, contributes one unit of angular momentum originating from the circular polarization of the laser.  In the case of monochromatic lasers [$g(\phi)=1$], these two numbers coincide ($s_0=n$) as given by Eq.~(\ref{S_fi_NCS_pl}).  

Next, we can further unveil the azimuth angle dependency associated with the initial (final) electron and the emitted photon spins by expanding the usual bispinor and polarization vector in terms of eigenfunctions of spin operators in the $z$ direction \cite{ivanov2022promises}:
\begin{eqnarray}
	\begin{aligned}
		u_{p,\lambda}&=\sum_{\sigma_{\lambda}=\pm\frac{1}{2}}e^{-i(\sigma_{\lambda}-\lambda)\varphi_{\bm{p}}}d_{\sigma_{\lambda},\lambda}^{1/2}(\theta_{\bm{p}})\begin{pmatrix} \sqrt{\varepsilon+m}\,w_{z}^{\sigma_{\lambda}} \\ 2\lambda\sqrt{\varepsilon-m}\,w_{z}^{\sigma_{\lambda}}    \\   \end{pmatrix},\\
		\bm{\epsilon}_{k,\Lambda}&=\sum_{\sigma_{\Lambda}=0,\pm1}e^{-i\sigma_{\Lambda}\varphi_{\bm{k}}}d_{\sigma_{\Lambda},\Lambda}^{1}(\theta_{\bm{k}})\bm{\chi}_{\sigma_{\Lambda}}\,,
	\end{aligned}
\end{eqnarray}
where $\theta_{\bm{p}}$ and $\theta_{\bm{k}}$  are the polar angles of the corresponding momentum vectors. The Wigner's $d$-functions are given by \cite{varshalovich1988quantum}
\begin{eqnarray}
	\begin{aligned}
		d_{\sigma,\lambda}^{1/2}(\theta)=\begin{pmatrix}
			\cos\theta/2 \qquad -\sin\theta/2 \\ \sin\theta/2 \qquad \cos\theta/2 \,
		\end{pmatrix}\,
	\end{aligned}
\end{eqnarray}
and
\begin{eqnarray}
	\begin{aligned}
		d_{\sigma_{\Lambda},\Lambda}^{1}(\theta)=\begin{pmatrix}
			\cos^2\theta/2 \qquad -\frac{1}{\sqrt{2}}\sin\theta \qquad \sin^2\theta/2 \\ 
			\frac{1}{\sqrt{2}}\sin\theta \qquad  \cos\theta  \qquad  -\frac{1}{\sqrt{2}}\sin\theta \\
			\sin^2\theta/2 \qquad \frac{1}{\sqrt{2}}\sin\theta \qquad \cos^2\theta/2
		\end{pmatrix}\,.
	\end{aligned}
\end{eqnarray}
{The spin basis functions satisfy
\begin{equation}
\hat{s}_{z} w^{\sigma_{\lambda}}_{z}=\sigma_{\lambda}\, w^{\sigma_{\lambda}}_{z}\,,\quad \hat{s}_{z}\bm{\chi}_{\sigma_{\Lambda}}=\sigma_{\Lambda} \bm{\chi}_{\sigma_{\Lambda}}\,,
\label{SAM_functions}
\end{equation}
with eigenvalues $\sigma_{\lambda}=\pm 1/2$ for the spinor and $\sigma_{\Lambda}=\pm 1, 0$ for the vector, respectively. The spinors are given by $ w^{1/2}_{z}=(1,0)^T$ and $ w^{-1/2}_{z}=(0,1)^T$, and the vectors are given by $\bm{\chi}_0=(0,0,1)^T$ and $\bm{\chi}_{\pm}=\frac{\mp}{\sqrt{2}}(1,\pm i,0)^T$.  
The corresponding spin operators in the $z$ direction $\hat{s}_z$ are a $2\times2$ Pauli matrix $\hat{s}_z=\frac{1}{2}\hat{\sigma}_z$ for the spinor and a $3\times3$ spin-1 matrix for the vector \cite{Bliokh:2015doa}. }

With the aforementioned considerations, we can now present the $S$-matrix element with the explicitly extracted azimuth angle dependency:
\begin{eqnarray}
\begin{aligned}
S_{fi}=&ie\frac{(2\pi)^{3}}{\omega}\delta^{3}(\bm{p}+s_{0}\bm{k}-\bm{p}^{\prime}-\bm{k}^{\prime})\sum_{n,\sigma,\ldots}e^{i(n-\sigma+\sigma_{\lambda^\prime}-\lambda^\prime)\varphi_{\bm{p}^{\prime}}}e^{i\sigma_{\Lambda^{\prime}}\varphi_{\bm{k}^\prime}}\,\\
&\times [\delta_{\sigma,0}(\delta_{\sigma_{\lambda^\prime},\sigma_{\lambda}}\mathcal{G}_{0}^{\uparrow\uparrow}+\delta_{\sigma_{\lambda^\prime},-\sigma_{\lambda}}\mathcal{G}_{0}^{\uparrow\downarrow})\\
&+(\delta_{\sigma_{\lambda^\prime},\sigma_{\lambda}}\mathcal{G}_{\pm}^{\uparrow\uparrow}+\delta_{\sigma_{\lambda^\prime},-\sigma_{\lambda}}\mathcal{G}_{\pm}^{\uparrow\downarrow})]\delta_{\sigma_\lambda,\lambda}d^{1/2}_{\sigma_{\lambda}^\prime,\lambda^\prime}(\theta_{\bm{p}^{\prime}})d^{1}_{\sigma_{\Lambda^{\prime}},\Lambda^\prime}(\theta_{\bm{k}^\prime}),
\end{aligned}
\label{S_NCS_final}
\end{eqnarray}
where the notations are as follows:
\begin{eqnarray}
	\begin{aligned}
		\mathcal{G}_{0}^{\uparrow\uparrow}=&\delta_{\sigma_{\Lambda^{\prime}},0}[2\sigma_{\lambda} f_{\lambda,\lambda^{\prime}}^{(2)}\,\mathscr{C}_{0}^{(n)}(s)\\&-2\Omega_{p}\Omega_{p^{\prime}}(f_{\lambda,\lambda^{\prime}}^{(1)}+2\sigma_{\lambda} f_{\lambda,\lambda^{\prime}}^{(2)})\,\mathscr{C}_{2}^{(n)}(s)]\,,\\
		\mathcal{G}_{0}^{\uparrow\downarrow}=&\delta_{\sigma_{\Lambda^{\prime}},2\sigma_{\lambda}}(-2\sigma_{\lambda}\sqrt{2})f_{\lambda,\lambda^{\prime}}^{(2)}\,\mathscr{C}_{0}^{(n)}(s)\,,\\
		\mathcal{G}_{\pm}^{\uparrow\uparrow}=&[\delta_{\sigma,2\sigma_{\lambda}}\delta_{\sigma_{\Lambda^{\prime}},2\sigma_{\lambda}}\sqrt{2}\Omega_{p^{\prime}}(2\sigma_\lambda f_{\lambda,\lambda^{\prime}}^{(1)}+f_{\lambda,\lambda^{\prime}}^{(2)})\,\\
		&-\delta_{\sigma,-2\sigma_{\lambda}}\delta_{\sigma_{\Lambda^{\prime}},-2\sigma_{\lambda}}\sqrt{2}\Omega_{p}(2\sigma_\lambda f_{\lambda,\lambda^{\prime}}^{(1)}+f_{\lambda,\lambda^{\prime}}^{(2)})]\,\mathscr{C}_{\sigma}^{(n)}(s)\,,\\
		\mathcal{G}_{\pm}^{\uparrow\downarrow}=&\delta_{\sigma,-2\sigma_{\lambda}}\delta_{\sigma_{\Lambda^{\prime}},0}[\Omega_{p^{\prime}}(-2\sigma_\lambda f_{\lambda,\lambda^{\prime}}^{(1)}+f_{\lambda,\lambda^{\prime}}^{(2)})\,\\
&+\Omega_{p}(2\sigma_\lambda f_{\lambda,\lambda^{\prime}}^{(1)}+f_{\lambda,\lambda^{\prime}}^{(2)})]\,\mathscr{C}_{\sigma}^{(n)}(s)\,,
	\end{aligned}
\end{eqnarray}
and 
\begin{eqnarray}
	\begin{aligned}
		f_{\lambda,\lambda^{\prime}}^{(1)}&=\sqrt{\varepsilon^{\prime}+m}\sqrt{\varepsilon+m}+2\lambda\,2\lambda^{\prime}\sqrt{\varepsilon^{\prime}-m}\sqrt{\varepsilon-m}\,,\\
		f_{\lambda,\lambda^{\prime}}^{(2)}&=2\lambda^{\prime}\sqrt{\varepsilon^{\prime}-m}\sqrt{\varepsilon+m}+2\lambda\sqrt{\varepsilon^{\prime}+m}\sqrt{\varepsilon-m}\,.\nonumber
	\end{aligned}
\end{eqnarray}
The summation in Eq.~(\ref{S_NCS_final}) can eliminate the dummy indices $\{\sigma,\sigma_{\lambda},\sigma_{\lambda^\prime},\sigma_{\Lambda^{\prime}}\}$ through the Kronecker deltas in the $S$-matrix element, see Appendix A. By further considering the relation $\varphi_{\bm{p}^{\prime}}=\varphi_{\bm{k}^{\prime}} \pm \pi$ supported by $\delta^{2}(\bm{p}_{\perp}^{\prime}+\bm{k}_{\perp}^{\prime})$, the azimuth-angle-dependent vortex phase of the $S$-matrix element can be revealed:  
\begin{eqnarray}
\begin{aligned}
S_{fi}=ie\frac{(2\pi)^{3}}{\omega}\delta^{3}(\bm{p}+s_{0}\bm{k}-\bm{p}^{\prime}-\bm{k}^{\prime})\sum_{n}e^{i(n+\lambda-\lambda^{\prime})\varphi_{\bm{p}^{\prime}}}\mathcal{M}^{(n)},\\
\end{aligned}
\label{S_NCS_vortex_phase}
\end{eqnarray}
where the harmonic amplitude $\mathcal{M}^{(n)}$ is given by Eq.~(\ref{a_amp_NCS}).  The phase factor $e^{i(n+\lambda-\lambda^{\prime})\varphi_{\bm{p}^{\prime}}}$ represents the complete phase structure of the NCS $S$-matrix element in the transverse plane, and the vortex structure can be inherited by the final particles, allowing them to possess OAM as discussed below.  

\subsection{Wave function of the radiated photon}
The outcome of a scattering process, also known as the evolved state, is related to the initial state via the $S$-matrix element \cite{Berestetskii1982quantum}:
\begin{eqnarray}
	\hat{S} \vert i \rangle=	\underset{f}{\int\mathllap{\sum}}\vert f\rangle \langle f \vert \hat{S} \vert i \rangle = \underset{f}{\int\mathllap{\sum}} \vert f\rangle \, S_{fi}\,,
\end{eqnarray}
where, in our case, the initial state $\vert i \rangle =\vert \bm{p},\lambda \rangle$ and the final state $ \vert f \rangle = \vert \bm{p}^\prime ,\lambda^\prime ; \bm{k}^\prime, \Lambda^ \prime\rangle = \vert \bm{p}^\prime ,\lambda^\prime \rangle \otimes \vert \bm{k}^\prime, \Lambda^ \prime\rangle$.  
Note that $\vert \bm{p},\lambda \rangle$, $\vert \bm{p}^\prime ,\lambda^\prime\rangle$ and $\vert \bm{k}^\prime, \Lambda^ \prime\rangle$ are the initial electron, the final electron, and the final photon eigenstates with  momentum and spin eigenvalues implied, respectively.  
After identifying the $S$-matrix element $S_{fi}=\langle p^{\prime},\lambda^{\prime} ; k^{\prime},\Lambda^{\prime}\vert \hat{S} \vert p,\lambda \rangle$, the evolved state of NCS can be expressed as
$\underset{\bm{p}^\prime ,\lambda^\prime , \bm{k}^\prime, \Lambda^ \prime}{\int\mathllap{\sum}} \vert \bm{p}^\prime ,\lambda^\prime \rangle \otimes \vert \bm{k}^\prime, \Lambda^ \prime\rangle S_{fi}$. 
Therefore, the $S$-matrix relates the final electron and radiated photon wave functions to each other, causing them to become entangled. Thus, a specific choice of wave packet for one of the final states directly affects the other \cite{ivanov2012creation,van2015inelastic,karlovets2022generation}.  
Moreover, the angular momentum transfer during the radiation process assumes the rotational symmetry in the transverse plane. Consequently, we project the final electron onto the following wave packet form: 
\begin{eqnarray}
	\int \frac{d^2{\bm{p}^{\prime}_{\perp}}}{(2\pi)^2} b_{\rho^\prime}(\bm{p}^\prime_\perp) \vert\bm{p}^\prime ,\lambda^\prime \rangle \,,
	\label{electron-WP}
\end{eqnarray}
where the Fourier amplitude for the wave packet $b_{\rho^\prime}(\bm{p}^\prime_\perp)=\sqrt{\frac{2\pi}{\rho^\prime}}\delta(p^\prime_\perp-\rho^\prime)$ fixes the transverse momentum $\vert \bm{p}^\prime_\perp\vert=\rho^\prime$ and integrates out the azimuth angle $\varphi_{\bm{p}^{\prime}}$.  The chosen wave packet, which is connected to the vortex amplitude in Eq. \eqref{vortex_amp} as $b_{\rho^\prime}(\bm{p}^\prime_\perp)=a_{\varkappa\,0}(\bm{p}^\prime_\perp)$, indicates that we are selecting the zero-vortex mode for the final electron state. However, projecting the outgoing electron onto the zero-vortex mode in the experimental setup presents a challenge. Nevertheless, as will be clear in the next subsection, with the wave packet choice in Eq.~\eqref{electron-WP} for the final electron, one obtains the same angular momentum transfer relation as in the classical radiation picture \cite{sasaki2008proposal,katoh2017angular}.  The resulting expression for the emitted photon wave function in momentum space is
\begin{eqnarray}
	\begin{aligned}
		\bm{A}_{\bm{k}^{\prime},\Lambda^{\prime}}&=\int \frac{d^2{\bm{p}^{\prime}_{\perp}}}{(2\pi)^2} b_{\rho^\prime}(\bm{p}^\prime_\perp)\,\bm{\epsilon}_{\bm{k}^{\prime},\Lambda^{\prime}}\, S_{fi}\,,\\
		&=ie\frac{2\pi}{\omega}\sqrt{\frac{2\pi}{k_{\perp}^{\prime}}}\delta(p_{z}+s_0k_{z}-p_{z}^{\prime}-k_{z}^{\prime})\delta(k_{\perp}^{\prime}-\rho^{\prime})\,\\
		&\times\sum_{n, \sigma_{\Lambda^{\prime}}}(-1)^{m_n^\prime}e^{i(m^\prime_n-\sigma_{\Lambda^{\prime}})\varphi_{\bm{k}^{\prime}}} d_{\sigma_{\Lambda^{\prime}},\Lambda^{\prime}}^{1}(\theta_{\bm{k}^{\prime}})\,\bm{\chi}_{\sigma_{\Lambda^{\prime}}}\mathcal{M}_{n}(s_0)\,,
	\end{aligned}
	\label{evolved-momentum}
\end{eqnarray}
where $S_{fi}$ is given in Eq.\eqref{S_NCS_vortex_phase}, and one could identify the Fourier amplitude of a Bessel vortex mode from the $\delta$-function $\delta(k_{\perp}^{\prime}-\rho^{\prime})$ and the spiral phase factor $e^{i(m^\prime_n-\sigma_{\Lambda^{\prime}})\varphi_{\bm{k}^{\prime}}}$ as in Eq.~(\ref{vortex_amp}).  
The position space wave function of the emitted photon is obtained via $\bm{A}_{\omega^{\prime},\Lambda^{\prime}}(\bm{x})=\int \frac{d^3{\bm{k}^{\prime}}}{(2\pi)^3} e^{i \bm{k}^\prime \bm{x}}  \bm{A}_{\bm{k}^{\prime},\Lambda^{\prime}}$, which gives
\begin{equation}
\begin{aligned}
\bm{A}_{\omega^{\prime},\Lambda^{\prime}}(\bm{x})&=ie\frac{1}{\omega}e^{i\,k_{z}^{\prime}z}\sqrt{\frac{k_\perp^\prime}{2\pi}}\sum_{n,\sigma_{\Lambda^{\prime}}}(-i)^{m^\prime_n+\sigma_{\Lambda^{\prime}}}d_{\sigma_{\Lambda^{\prime}},\Lambda^{\prime}}^{1}(\theta_{\bm{k}^{\prime}})\,\\
&\times e^{i(m^\prime_n-\sigma_{\Lambda^{\prime}})\varphi_{\bm{r}}}J_{m^\prime_n-\sigma_{\Lambda^{\prime}}}(k^{\prime}_\perp r)\,\bm{\chi}_{\sigma_{\Lambda^{\prime}}}\,\mathcal{M}_{n}(s_0) \,,
\end{aligned}
\label{vortex_state_config}
\end{equation}
where $m^\prime_n=n+\lambda-\lambda^{\prime}$ is inherited from the NCS $S$-matrix element in Eq.~\eqref{S_NCS_vortex_phase}.
By applying the total angular momentum (TAM) operator $\hat{J}_z=\hat{s}_z+\hat{L}_z$ to Eq.~\eqref{vortex_state_config}, it is found that $\hat{J}\bm{A}_{\omega^{\prime},\Lambda^{\prime}}(\bm{x})=m^\prime_n~\bm{A}_{\omega^{\prime},\Lambda^{\prime}}(\bm{x})$. This implies that $m^\prime_n$ represents the TAM carried by the vortex photon \cite{knyazev2018beams}. 

The phase factor $e^{i(m^\prime_n-\sigma_{\Lambda^{\prime}})\varphi_{\bm{r}}}$ encodes two kinds of spin-orbit interactions (SOI): the intrinsic SOI of vortex light \cite{bliokh2015nature} determining its OAM number $l^\prime_n$ with three values $\lbrace m^\prime_n-\Lambda^{\prime},m^\prime_n, m^\prime_n +\Lambda^{\prime}\rbrace$ \cite{jentschura2011generation}, and the SOI of the radiation dynamics where the SAM of electrons and laser photons contribute to the emitted photon's OAM. For an incoming ultrarelativistic electron, the cone angle associated with the radiation is estimated as $\theta_{\bm{k}^{\prime}}\sim\frac{1}{\gamma_0}\ll1$ (where $\gamma_0=\frac{\varepsilon}{m_e}$ is the electron's relativistic factor), and the Wigner function simplifies to a Kronecker delta $d_{\sigma_{\Lambda^{\prime}},\Lambda^{\prime}}^{1}(\theta_{\bm{k}^{\prime}})\approx\delta_{\sigma_{\Lambda^{\prime}},\Lambda^{\prime}}$. This allows for interpreting the wave function of the radiated $\gamma$ photon as the eigenfunction of SAM and OAM operators with respective eigenvalues $\Lambda^\prime$ and $l^\prime_n=m_n^\prime-\Lambda^{\prime}$, which is similar in optics where the TAM of a paraxial vortex light splits into SAM and OAM \cite{Leader:2015vwa,Leader:2017htb}. Moreover, for the radiation in moderate laser intensities ($\xi \sim 1$), the electron spin-flip channel ($\lambda\ne\lambda^{\prime}$) is negligible, resulting in $m^\prime_n=n$. Consequently, the $\gamma$ photon emitted from the $n$th harmonic of the NCS corresponds to the vortex mode with SAM $\Lambda^{\prime}$ and OAM $l^\prime_n=n-\Lambda^{\prime}$.

The relation $m^\prime_n=n$ suggests that SAM of $n$ laser photons goes completely into the OAM of the emitted single $\gamma$ photon, which is a consequence of assigning zero OAM to the final electron. In principle, the SAM of multiple laser photons could contribute to both the final electron and the $\gamma$ photon, provided that a coherent projection of the final electron onto a vortex mode is performed. However, it should be noted that, currently, there is no such control over the final electron in the experiment. 
The wave function of the $\gamma$ photon is related to the electron via the $S$-matrix element  \cite{Berestetskii1982quantum,ivanov2012creation,van2015inelastic}, and identifying the $\gamma$ photon as a vortex mode requires extra conditions over the final electron, which are missing in classical theories. 

\subsection{Probability of radiation of vortex $\gamma$ photons}
To obtain the probability for vortex photon radiation, we project the evolved photon state from Eq.~\eqref{evolved-momentum} onto a basis of OAM-sensitive detectors.  This allows us to relate the $S$-matrix for the generation of the vortex photon to the usual plane wave NCS matrix element $S_{fi}$ obtained in Eq.~\eqref{S_NCS_final}:
\begin{eqnarray}
	\begin{aligned}
		S^{\text{vortex}}_{fi}&=\int\frac{d^{2}\bm{k}_\perp^{\prime}}{(2\pi)^{2}}\,a^{\ast}_{{\kappa}^{\prime} m_{\kappa^\prime}}(\bm{k}_\perp^{\prime})\,\int \frac{d^2{\bm{p}^{\prime}_{\perp}}}{(2\pi)^2} b_{\rho^\prime}(\bm{p}^\prime_\perp)\,S_{fi}\,,\\
		&=\frac{ie 2\pi }{\omega}\delta(p_{z}+s_{0}k_{z}-p^{\prime}_{z}-k^{\prime}_{z})\delta({\rho}^{\prime}-\kappa^{\prime})\,\\
		& \times\sum_{n}\,i^{-m_{\kappa^\prime}}\,\delta_{m_{\kappa^\prime},\,n+\lambda-\lambda^{\prime}}\mathcal{M}_{n}(s_{0})\,,
	\end{aligned}
	\label{s-vortex}
\end{eqnarray}
where the Fourier amplitudes $a^{\ast}_{{\kappa}^{\prime} m_{\kappa^\prime}}(\bm{k}_\perp^{\prime})$ and $b_{\rho^\prime}(\bm{p}^\prime_\perp)$ correspond to the projection of the final photon and electron onto the vortex and zero-vortex modes, respectively.
After employing the normalization convention for final state particle wave packets as in \cite{jentschura2011compton}, and including the factors $\sqrt{\frac{1}{2\varepsilon V}}$, $\sqrt{\frac{\pi}{2\varepsilon^\prime RL_z}}$, and $\sqrt{\frac{\pi}{2\omega^\prime RLz}}$ in the corresponding wave functions, we can write the probability for the vortex photon emission as
\begin{eqnarray}
dW^{\text{vortex}}=|S^{\text{vortex}}_{fi}|^{2}\frac{RL_zdp_{\perp}^{\prime}dp_{z}^{\prime}}{2\pi^{2}}\frac{RL_zdk_{\perp}^{\prime}dk_{z}^{\prime}}{2\pi^{2}}.
\label{rate}
\end{eqnarray}
The differential rate for the generation of the vortex photon with TAM $m_{\kappa^\prime}$ is
\begin{eqnarray}
\begin{aligned}
\frac{d^2W^{\text{vortex}}}{d\omega^\prime d\theta_{\bm{k}^\prime} }=\frac{e^2}{4\pi}\frac{k^\prime_\perp}{(kp)(kp^{\prime})}\sum_{n}\delta_{m_{\kappa^\prime},\,n+\lambda-\lambda^{\prime}}|\mathcal{M}_{n}(s_{0})|^{2}\,,
\end{aligned}\label{rate_vortex}
\end{eqnarray}
where a factor $\frac{1}{\pi R}$ from the normalization is replaced by $k^\prime_\perp=\omega^\prime\sin\theta_{\bm{k}^\prime}$. Here, the angular momentum conservation imposed by the $\delta$-function $m_{\kappa^\prime}=n+\lambda-\lambda^\prime$ indicates that TAM of the radiated $\gamma$ photon $m_{\kappa^\prime}$ receives contributions from the laser photon SAM $n$ and electron spin flip $(\lambda-\lambda^\prime)$. If we neglect the spin-flip channel ($\lambda \ne \lambda^{\prime}$) of NCS for moderate laser intensities ($\xi \sim 1$), we get $m_{\kappa^\prime}=n$, which is consistent with the classical results of the radiation from the spiral motion of a charged point particle \cite{sasaki2008proposal,bahrdt2013first,katoh2017angular}.  Note that in Eq.~(\ref{rate}), there is a sum over harmonics inside the square, and one should write $\vert \sum_n S_{fi}^{(n)}\vert^2=\vert\sum_{n=n^\prime} S_{fi}^{(n)}S_{fi}^{\ast(n^\prime)}\vert+\vert\sum_{n \neq n^\prime} S_{fi}^{(n)}S_{fi}^{\ast(n^\prime)}\vert$. The interference terms $\vert\sum_{n \neq n^\prime} S_{fi}^{(n)}S_{fi}^{\ast(n^\prime)}\vert$ arise due to the finite duration of the laser, which causes the ponderomotive broadening and, in turn, results in the existence of different harmonics for the same radiation kinematics $s_0<n\le s_0-\Delta\beta$ \cite{seipt2016analytical}. 
However, the interference terms are eliminated by the Kronecker delta resulting from angular momentum conservation in Eq.~(\ref{rate_vortex}). 
In the case of a monochromatic laser, the interference terms are canceled out by the energy-momentum conserving delta function, which contains the discrete harmonic label.

\section{Numerical Results}
\subsection{Radiation rate of the vortex $\gamma$ photon}

\begin{figure}
	\setlength{\abovecaptionskip}{-0.0cm}
	\includegraphics[width=.95\linewidth]{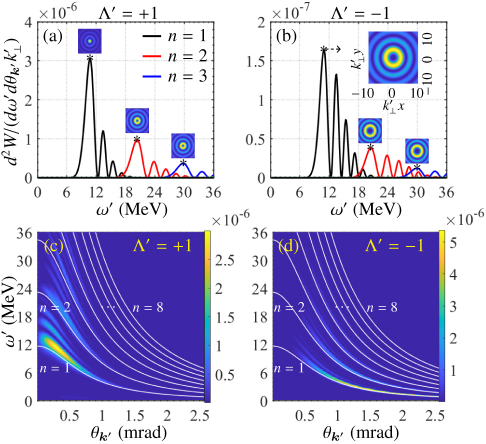}
	\begin{picture}(300,0)           
	\end{picture}
	\caption{Radiation rates for the vortex $\gamma$ photons with $\theta_{\bm{k}^{\prime}}\approx0.35$ mrad for the first three harmonics for helicities (a) $\Lambda^{\prime}= +1$ and (b) $\Lambda^{\prime}=- 1$. Inset: transverse intensity distributions given in arbitrary units calculated from $\vert \bm{A}_{\omega^{\prime}}(\bm{x})\vert ^2$ for $\omega^\prime$ taken at the main harmonic peaks marked by asterisks.  Angle-resolved energy spectra in Eq.~\eqref{rate_vortex} for photon helicities (c) $\Lambda^{\prime}= +1$ and (d) $\Lambda^{\prime}=- 1$; white lines stand for kinematic relations for harmonics $\omega^\prime_{n}(\theta_{\bm{k}^{\prime}})$ in the monochromatic laser.  }
	\label{fig_coherent}
\end{figure}

The analytical results in the previous section hold for optical lasers with the pulse duration $\tau \gtrsim 20$ fs regardless of the electron energy and the laser intensity as long as the single photon emission along the collision axis is dominated.  
In the following, we present numerical results where we assume the electron energy 
 $\varepsilon=1$ GeV, the laser intensity $\xi=1$, the central laser photon energy $\omega=1.55$ eV, and the pulse shape $g(\phi)=\cos^{2}(\frac{\phi}{2N_\text{cycle}})$ with $N_\text{cycle}=10$ (corresponding to an optical laser with $\lambda_0\approx0.8\,\mu\text{m}$, $I\approx2\times10^{18}$ W/cm$^{2}$, and  $\tau\approx26.7$ fs). The results for monochromatic lasers are presented in Appendix B. Figure~\ref{fig_coherent} shows the radiation rates of polarized vortex $\gamma$ photons calculated via Eq.~\eqref{rate_vortex}.  We present the energy spectra for a fixed angle in Figs.~\ref{fig_coherent} (a) and \ref{fig_coherent} (b).    
 The $n$th harmonic corresponds to the vortex state with TAM number $m^\prime_n=n$ and OAM number $l^\prime_n=n-\Lambda^\prime$.  The transversal distributions for $\vert\bm{A}_{\omega^{\prime}}(\bm{x})\vert^2$ in the inset feature the doughnut shape of the photon with nonzero OAM.  Moreover, the ponderomotive broadening due to the finite pulse shape would cause the nearby harmonics to overlap \cite{seipt2016analytical}, and one could find multiple OAM modes for certain energy and polarization.  This feature holds for a wide range of $\theta_{\bm{k}^{\prime}}$ as seen from the angle-resolved spectra given in Figs.~\ref{fig_coherent} (c) and \ref{fig_coherent} (d), where the white lines stand for the kinematic relation for the monochromatic laser.  Each group of coordinates $(\Lambda^\prime,\omega^\prime,\theta_{\bm{k}^{\prime}})$ represents an independent radiation event and defines a coherent vortex state with fixed OAM.

\begin{figure}
	\setlength{\abovecaptionskip}{-0.0cm}
	\includegraphics[width=1.0\linewidth]{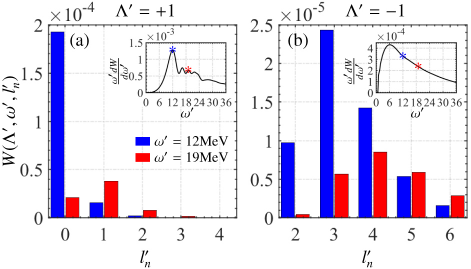}
	\begin{picture}(300,0)
	\end{picture}
	\caption{ OAM distributions of incoherent vortex $\gamma$ photons with energies $\omega^\prime=12$ MeV and $\omega^\prime=19$ MeV for helicities (a) $\Lambda^{\prime}= +1$ and (b) $\Lambda^{\prime}=- 1$. Inset: power spectra of the emitted polarized photons $\frac{\omega^\prime d W}{d\omega^\prime}$ where two energies are marked by asterisks with the same colors as in the OAM distributions. 	Note that OAM values start from $l_1^\prime=0$ for $\Lambda^\prime=+1$ and $l_1^\prime=2$ for $\Lambda^\prime=-1$ because of $l^\prime_n=n-\Lambda^\prime$.}
	\label{fig_incoherent}
\end{figure} 

Since vortex photons formed from radiation with different $\theta_{\bm{k}^{\prime}}$ overlap in the transverse position space, we introduce an incoherent summation over $\theta_{\bm{k}^{\prime}}$ in Eq.~\eqref{rate_vortex} and obtain $\frac{dW_\text{vortex}}{d\omega^\prime\,k_\perp^\prime}\equiv\sum_n W(\Lambda^\prime,\omega^\prime,l_n^\prime)$, where we have shifted from $m_n^\prime$ to $l_n^\prime$ for the convenience of discussing OAM.  
We then observe that $\gamma$ photons with definite energy and polarization $(\Lambda^\prime,\omega^\prime)$ (marked by two asterisks in the inset of Fig.~\ref{fig_incoherent}) can be identified as a mixed state of vortex $\gamma$ photons with different OAM numbers (see OAM distributions in Fig.~\ref{fig_incoherent}).  This could be explained as follows.
For the same energy, higher harmonics also contribute, with an increasing $\theta_{\bm{k}}$ in the summation, bringing with them a large amount of OAM.  In addition, if we define the polarization degree for each harmonic as $P_n(\omega^\prime)=\frac{ W(\Lambda^\prime=+1,\,\omega^\prime,\,n)-W(\Lambda^\prime=-1,\,\omega^\prime,\,n)}{\vert W(\Lambda^\prime=+1,\,\omega^\prime,\,n)+W(\Lambda^\prime=-1,\,\omega^\prime,\,n)\vert}$, we get, for the polarized vortex $\gamma$ photon with $\omega^\prime=12$ MeV, $P_1\approx90.37$\%, $P_2\approx-23.44$\%, $P_3\approx-74.16$\%, etc., which are different from the results for the plane wave $\gamma$ photon $P(\omega^\prime=12$MeV$)\approx58.95$\% (in the latter, all harmonics are counted).  
We stress that such an incoherent nature of the generated vortex photon is the crucial feature of strong laser-based $\gamma$-ray sources.  
This new recognition of vortex $\gamma$ photons in turn calls for novel considerations with regard to its applications in nuclear or particle physics studies, in which the current assumption is that the available vortex particles are in a pure coherent state \cite{jentschura2011generation,ivanov2020doing,ivanov2022promises}.  

\subsection{Beam properties of the vortex $\gamma$ photon}

\begin{figure}
	\setlength{\abovecaptionskip}{-0.0cm}
	\includegraphics[width=1\linewidth]{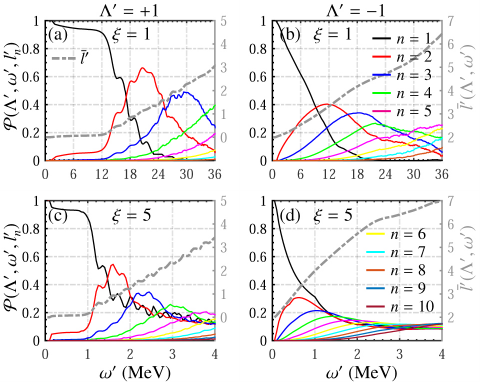}
	\begin{picture}(300,0)  
	\end{picture}
	\caption{OAM ratios $\mathcal{P}(\Lambda^\prime, \omega^\prime,l^\prime_n)$ over energy spectra for polarized vortex $\gamma$ photons. Results are presented up to the tenth harmonic (solid lines) and corresponding OAM values are implied via $l^\prime_n=n-\Lambda^\prime$.  The average OAM number $\bar{l}^\prime(\Lambda^\prime, \omega^\prime)$ is given by the gray dashed line.
	}
	\label{fig_oam-ratio}
\end{figure}

To analyze the property of the vortex $\gamma$ photon beam over the whole energy range, we introduce OAM ratios of polarized vortex photons in the beam as $\mathcal{P}(\Lambda^\prime,\omega^\prime,l^\prime_n)=\frac{W(\Lambda^\prime,\omega^\prime,l^\prime_n)}{\sum_{n} {W(\Lambda^\prime, \omega^\prime,l^\prime_n)}}$.
Corresponding numerical results are given in Fig.~\ref{fig_oam-ratio} for laser intensities $\xi=1$ and $\xi=5$  ($I\approx 5\times10^{19}$ W/cm$^{2}$), and other parameters are the same as those in Fig.~\ref{fig_coherent}.  
At low energy regimes below the Compton edge, the first harmonic dominates the radiation; thus, the corresponding $\gamma$ photons carry nonzero OAM only for $\Lambda^\prime=-1$ due to the relation $l^\prime_1=1-\Lambda^\prime$. As the energy increases, the appearance of the peak in each harmonic indicates the dominance of the corresponding OAM mode [see Figs.~\ref{fig_oam-ratio} (a) and ~\ref{fig_oam-ratio} (b)]. However, when the laser intensity increases, the feature is only weakly preserved for $\Lambda^\prime=+1$ [see Fig.~\ref{fig_oam-ratio} (c)].  
The average OAM $\bar{l}^\prime(\Lambda^\prime,\omega^\prime)=\sum_{n} \mathcal{P}(\Lambda^\prime,\omega^\prime,l^\prime_n)\,l^\prime_n$ carried by the $\gamma$ photon gradually increases with energy, since more harmonics contribute, while the lower harmonics is also present due to the ponderomotive broadening caused by the finite laser pulse. Our results indicate that the $\gamma$ photons radiated in a CP laser could possess intrinsic OAM inherited from SAM of absorbed laser photons. This is missing in previous studies \cite{liu2016generation,Chen:2018tkb,gong2018brilliant,liu2020vortex,wang2020generation,hu2021attosecond,zhang2021efficient,bake2022bright}, in which SAM and OAM of the laser photons transfer wholly to the mechanical OAM of the final $\gamma$ photon and electron.  Such differences in understanding the angular momentum transfer relation in NCS need to be resolved, especially when the single photon emission is dominated as is assumed in this study. 
These are most likely relevant for the intermediate regime ($0.1\lesssim\xi\lesssim10$) which is currently under investigation \cite{di2019improved,blackburn2021local} and available for future experiments \cite{abramowicz2021conceptual}.

\section{Conclusion}

Regarding the experimental feasibility of achieving spin-to-orbital angular momentum transfer in NCS, one may try to relax some of the theoretical assumptions.  For example, even the slightest deviation of the incoming electron from the collision axis would cause significant OAM broadening of the $\gamma$ photon.  However, an effective angular momentum transfer relation $m^\prime_\gamma \approx n$ can still be achieved by considering the off-axis emission of the vortex $\gamma$ photon.  Moreover, the infinite coherence length implied by the plane wave state of the incoming electron could also be replaced with a finite one.  The most challenging requirement of projecting the final state electron onto a vortex state with zero OAM can also be eliminated if we prepare the initial electron with transverse coherence and consider the off axis emission of the vortex $\gamma$ photon. However, in these instances, the vortex $\gamma$ photons are emitted off-axis and could potentially be in ``spatiotemporal vortex" modes \cite{Bliokh:2012az}. A detailed discussion of these scenarios requires additional study. 

So far, it is clear that the spin-to-orbital angular momentum transfer in NCS is a crucial point concerning the generation of vortex $\gamma$ photons. Although studies based on the  classical radiation theory predict the OAM of the vortex $\gamma$ photon in CP lasers \cite{taira2017gamma}, quantum treatment becomes necessary to grasp the transverse coherence condition on the electron for the $\gamma$ photon to possess a vortex mode.   

In summary, we investigate the generation of vortex $\gamma$ photons in a CP laser pulse by employing a vortex scattering theory in the Furry picture of strong-field QED.  
We show that, when the final electron beam satisfies a transverse coherence condition, $\gamma$ photons corresponding to NCS harmonics can be identified by vortex modes with intrinsic OAM as a result of spin-to-orbital angular momentum transfer during multiphoton absorption. However, due to the intense laser, the generated $\gamma$ photons are, in general, in the incoherent mixed state of various vortex modes.  
Furthermore, the attained OAM is determined by the laser photon absorption number, and high OAM $\gamma$ photons could be achieved by utilizing OAM of vortex state electrons \cite{lloyd2017electron,bliokh2017theory} and/or vortex lasers.  
Finally, vortex $\gamma$ photons may unveil new possibilities in strong-field QED studies, besides their already envisioned applications in nuclear and particle physics related research.  \\

{\it Acknowledgment:}  We thank C. H. Keitel, K. Z. Hatsagortsyan, P. Zhang, and I. P. Ivanov for helpful discussions. We are grateful to Q. Zhao for his help during the revision of our manuscript. This work is supported by the National Natural Science Foundation of China (Grants No. 11874295, No. 12022506, No. U2267204, No. 11905169, No. 12275209, No. 12147176, No. 12135001, No. 11825502, No. 11921006, and No. U2241281), the Foundation of Science and Technology on Plasma Physics Laboratory (No. JCKYS2021212008), and the Shaanxi Fundamental Science Research Project for Mathematics and Physics (Grant No. 22JSY014).
 
\appendix
\section{VORTEX PHASE OF THE $S$-MATRIX ELEMENT}
The final form of the $S$-matrix element with the vortex phase explicitly extracted reads
\begin{eqnarray}
	S_{fi}=ie\frac{(2\pi)^{3}}{\omega}\delta^{3}(\bm{p}+s_{0}\bm{k}-\bm{p}^{\prime}-\bm{k}^{\prime})\sum_{n}e^{i(n+\lambda-\lambda^{\prime})\varphi_{\bm{p}^{\prime}}}\mathcal{M}^{(n)}(s_0),\nonumber
	\label{a_S_NCS_final}
\end{eqnarray}
where the amplitude is given by
\begin{eqnarray}
\begin{aligned}
		\mathcal{M}^{(n)}(s)&=d_{\lambda,\lambda^{\prime}}^{1/2}(\theta_{\bm{p}^{\prime}})\,d_{0,\Lambda^{\prime}}^{1}(\theta_{\bm{k}^{\prime}}) \,\tilde{\mathcal{G}}_{0}^{\uparrow\uparrow}(s)\\
		&+e^{-i2\lambda( \varphi_{\bm{p}^{\prime}}-\varphi_{\bm{k}^{\prime}})}d_{-\lambda,\lambda^{\prime}}^{1/2}(\theta_{\bm{p}^{\prime}})\,d_{2\lambda,\Lambda^{\prime}}^{1}(\theta_{\bm{k}^{\prime}})  \,\tilde{\mathcal{G}}_{0}^{\uparrow\downarrow}(s)\\
		&+e^{-i2\lambda( \varphi_{\bm{p}^{\prime}}-\varphi_{\bm{k}^{\prime}})}d_{\lambda,\lambda^{\prime}}^{1/2}(\theta_{\bm{p}^{\prime}})\,d_{2\lambda,\Lambda^{\prime}}^{1}(\theta_{\bm{k}^{\prime}})  \,\tilde{\mathcal{G}}_{\pm,1}^{\uparrow\uparrow}(s)\\
		&+e^{i2\lambda( \varphi_{\bm{p}^{\prime}}-\varphi_{\bm{k}^{\prime}})}d_{\lambda,\lambda^{\prime}}^{1/2}(\theta_{\bm{p}^{\prime}})\,d_{-2\lambda,\Lambda^{\prime}}^{1}(\theta_{\bm{k}^{\prime}})  \,\tilde{\mathcal{G}}_{\pm,2}^{\uparrow\uparrow}(s)\\
		&+d_{-\lambda,\lambda^{\prime}}^{1/2}(\theta_{\bm{p}^{\prime}})\,d_{0,\Lambda^{\prime}}^{1}(\theta_{\bm{k}^{\prime}})  \,\tilde{\mathcal{G}}_{\pm}^{\uparrow\downarrow}(s) \,,
\end{aligned}
\label{a_amp_NCS}
\end{eqnarray}
with the following notations:
\begin{eqnarray}
	\begin{aligned}
		\tilde{\mathcal{G}}_{0}^{\uparrow\uparrow}(s)&=2\lambda f_{\lambda,\lambda^{\prime}}^{(2)}\,\mathscr{C}_{0}^{(n)}(s)-2\Omega_{p}\Omega_{p^{\prime}}(f_{\lambda,\lambda^{\prime}}^{(1)}+2\lambda f_{\lambda,\lambda^{\prime}}^{(2)})\,\mathscr{C}_{2}^{(n)}(s)\,,\nonumber\\
		\tilde{\mathcal{G}}_{0}^{\uparrow\downarrow}(s)&=-2\sqrt{2}\lambda f_{\lambda,\lambda^{\prime}}^{(2)}\,\mathscr{C}_{0}^{(n)}(s)\,,\nonumber\\
		\tilde{\mathcal{G}}_{\pm,1}^{\uparrow\uparrow}(s)&=\sqrt{2}\Omega_{p^{\prime}}(2\lambda f_{\lambda,\lambda^{\prime}}^{(1)}+f_{\lambda,\lambda^{\prime}}^{(2)})\,\mathscr{C}_{2\lambda}^{(n)}(s)\,,\nonumber\\
		\tilde{\mathcal{G}}_{\pm,2}^{\uparrow\uparrow}(s)&=-\sqrt{2}\Omega_{p}(2\lambda f_{\lambda,\lambda^{\prime}}^{(1)}+f_{\lambda,\lambda^{\prime}}^{(2)})\,\mathscr{C}_{-2\lambda}^{(n)}(s)\,,\nonumber\\
		\tilde{\mathcal{G}}_{\pm}^{\uparrow\downarrow}(s)&=[\Omega_{p^{\prime}}(-2\lambda f_{\lambda,\lambda^{\prime}}^{(1)}+f_{\lambda,\lambda^{\prime}}^{(2)})+\Omega_{p}(2\lambda f_{\lambda,\lambda^{\prime}}^{(1)}+f_{\lambda,\lambda^{\prime}}^{(2)})]\,\mathscr{C}_{-2\lambda}^{(n)}(s)\,.
	\end{aligned}
\end{eqnarray}
Note that in Eq.\eqref{a_amp_NCS}, the azimuthal angle dependency will be eliminated by considering the relation $\varphi_{\bm{p}^{\prime}}=\varphi_{\bm{k}^{\prime}} \pm \pi$  supported by $\delta^{2}(\bm{p}_{\perp}^{\prime}+\bm{k}_{\perp}^{\prime})$.  Therefore, the complex phase factor $e^{i(n+\lambda-\lambda^{\prime})\varphi_{\bm{p}^{\prime}}}$ holds the only remaining azimuth angle dependency of the $S$-matrix element.

We can also write a more generic form of the $S$-matrix element with azimuth-angle-dependent phase explicitly extracted:
\begin{eqnarray}
	\begin{aligned}
		S_{fi}=&ie\frac{(2\pi)^{3}}{\omega}\delta^{3}(\bm{p}+s_{0}\bm{k}-\bm{p}^{\prime}-\bm{k}^{\prime})\sum_{n_1,n_2,\sigma,\ldots}\,e^{-i\tilde{m}_{\bm{p}}\varphi_{\bm{p}}}\,\\
		&\times e^{i\tilde{m}_{\bm{p}^\prime}\varphi_{\bm{p}^{\prime}}}e^{i\tilde{m}_{\bm{k}^\prime}\varphi_{\bm{k}^\prime}}
		 d^{1/2}_{\sigma_{\lambda},\lambda}(\theta_{\bm{p}})d^{1/2}_{\sigma_{\lambda}^\prime,\lambda^\prime}(\theta_{\bm{p}^{\prime}})d^{1}_{\sigma_{\Lambda^{\prime}},\Lambda^\prime}(\theta_{\bm{k}^\prime})\\
		&\times[ \delta_{\sigma,0}(\delta_{\sigma_{\lambda^\prime},\sigma_{\lambda}}\mathcal{G}_{0}^{\uparrow\uparrow(n_1,n_2)}+\delta_{\sigma_{\lambda^\prime},-\sigma_{\lambda}}\mathcal{G}_{0}^{\uparrow\downarrow(n_1,n_2)})\\&+(\delta_{\sigma_{\lambda^\prime},\sigma_{\lambda}}\mathcal{G}_{\pm}^{\uparrow\uparrow(n_1,n_2)}+\delta_{\sigma_{\lambda^\prime},-\sigma_{\lambda}}\mathcal{G}_{\pm}^{\uparrow\downarrow(n_1,n_2)})],
	\end{aligned}
	\label{a_S_NCS_final}
\end{eqnarray} 
where we have introduced the notations $\tilde{m}_{\bm{p}}=n_1+\sigma_{\lambda}-\lambda$, $\tilde{m}_{\bm{p}^\prime}=n_2-\sigma+\sigma_{\lambda^\prime}-\lambda^\prime$, and $\tilde{m}_{\bm{k}^\prime}=\sigma_{\Lambda^{\prime}}$. Comparing this with Eq.~(\ref{S_NCS_final}), it is evident that Eq.~(\ref{a_S_NCS_final}) also includes the contribution from the initial electron. In the main text, it is assumed that the initial electron propagates along the $z$ axis and has zero transverse momentum ($\theta_{\bm{p}}=0$). Therefore, the initial electron does not contribute to the azimuth angle phase in Eq.~(\ref{S_NCS_final}).  Moreover, in Eq.~(\ref{a_S_NCS_final}), there are two harmonic labels present. This introduces an additional factor to $\mathcal{G}$ through the integral definitions of $\mathscr{C}_{\sigma}^{(n_1,n_2)}=\mathscr{C}_{\sigma}^{(n_2)}e^{in_1\phi}J_{n_1}(\alpha_{p_{\perp}}g(\phi))$, where $\mathscr{C}_{\sigma}^{(n_2)}$ is defined in the main text. Using the expression in Eq.~(\ref{a_S_NCS_final}), it becomes possible to calculate various physics scenarios related to vortex photon radiation during NCS processes. For example, one can study triple vortex scattering or consider the case in which the initial electron possesses transverse coherency.   

\section{VORTEX $\gamma$ PHOTON RADIATION IN MONOCHROMATIC LASERS}
For the monochromatic laser case, after performing the harmonic expansion and using SAM eigenfunctions for electron and photon states, the $S$-matrix element is written as
\begin{equation}
	S_{fi}
	=\mathcal{N}^{\prime}_\text{{NCS}} \sum_n \delta^{(4)}(q + n k- q^{\prime}-{k}^{\prime})\,e^{i( n+\lambda-\lambda^{\prime})\varphi_{\bm{p}^{\prime}}} {M}_{n},	
	\label{S_fi_NCS_pl}
\end{equation}	 
where $\mathcal{N}^{\prime}_\text{{NCS}}=ie(2\pi)^4$ and the amplitude reads 
\begin{eqnarray}
	\begin{aligned}
		{M}_{n}&=d_{\lambda,\lambda^{\prime}}^{1/2}(\theta_{\bm{p}^{\prime}})\,d_{0,\Lambda^{\prime}}^{1}(\theta_{\bm{k}^{\prime}}) \,{G}_{0}^{\uparrow\uparrow}\\
		&+e^{-i2\lambda( \varphi_{\bm{p}^{\prime}}-\varphi_{\bm{k}^{\prime}})}d_{-\lambda,\lambda^{\prime}}^{1/2}(\theta_{\bm{p}^{\prime}})\,d_{2\lambda,\Lambda^{\prime}}^{1}(\theta_{\bm{k}^{\prime}})  \,{G}_{0}^{\uparrow\downarrow}\\
		&+e^{-i2\lambda( \varphi_{\bm{p}^{\prime}}-\varphi_{\bm{k}^{\prime}})}d_{\lambda,\lambda^{\prime}}^{1/2}(\theta_{\bm{p}^{\prime}})\,d_{2\lambda,\Lambda^{\prime}}^{1}(\theta_{\bm{k}^{\prime}})  \,{G}_{\pm,1}^{\uparrow\uparrow}\\
		&+e^{i2\lambda( \varphi_{\bm{p}^{\prime}}-\varphi_{\bm{k}^{\prime}})}d_{\lambda,\lambda^{\prime}}^{1/2}(\theta_{\bm{p}^{\prime}})\,d_{-2\lambda,\Lambda^{\prime}}^{1}(\theta_{\bm{k}^{\prime}})  \,{G}_{\pm,2}^{\uparrow\uparrow}\\
		&+d_{-\lambda,\lambda^{\prime}}^{1/2}(\theta_{\bm{p}^{\prime}})\,d_{0,\Lambda^{\prime}}^{1}(\theta_{\bm{k}^{\prime}})  \,{G}_{\pm}^{\uparrow\downarrow} \,,
	\end{aligned}
	\label{amp_NCS_pl}
\end{eqnarray}
with the following notations:
\begin{eqnarray}
	\begin{aligned}
		{G}_{0}^{\uparrow\uparrow}&=2\lambda f_{\lambda,\lambda^{\prime}}^{(2)}\,J_n(\alpha_{p_{\perp}^{\prime}})-2\beta_{p}\beta_{p^{\prime}}(f_{\lambda,\lambda^{\prime}}^{(1)}+2\lambda f_{\lambda,\lambda^{\prime}}^{(2)})\,J_n(\alpha_{p_{\perp}^{\prime}})\,,\nonumber\\
		{G}_{0}^{\uparrow\downarrow}&=-2\sqrt{2}\lambda f_{\lambda,\lambda^{\prime}}^{(2)}\,J_n(\alpha_{p_{\perp}^{\prime}})\,,\nonumber\\
		{G}_{\pm,1}^{\uparrow\uparrow}&=\sqrt{2}\beta_{p^{\prime}}(2\lambda f_{\lambda,\lambda^{\prime}}^{(1)}+f_{\lambda,\lambda^{\prime}}^{(2)})\,J_{n-2\lambda}(\alpha_{p_{\perp}^{\prime}})\,,\nonumber\\
		{G}_{\pm,2}^{\uparrow\uparrow}&=-\sqrt{2}\beta_{p}(2\lambda f_{\lambda,\lambda^{\prime}}^{(1)}+f_{\lambda,\lambda^{\prime}}^{(2)})\,J_{n+2\lambda}(\alpha_{p_{\perp}^{\prime}})\,,\nonumber\\
		{G}_{\pm}^{\uparrow\downarrow}&=[\beta_{p^{\prime}}(-2\lambda f_{\lambda,\lambda^{\prime}}^{(1)}+f_{\lambda,\lambda^{\prime}}^{(2)})-\beta_{p}(-2\lambda f_{\lambda,\lambda^{\prime}}^{(1)}-f_{\lambda,\lambda^{\prime}}^{(2)})]\,J_{n+2\lambda}(\alpha_{p_{\perp}^{\prime}})\,.
	\end{aligned}
\end{eqnarray}  
Note that the full expression of the azimuth-angle-dependent phase in the transverse plane is needed to explore the vortex mode of the NCS $S$-matrix element. If we perform harmonic expansion alone, as is done in textbooks \cite{Berestetskii1982quantum}, we only get the phase factor $e^{i  n \varphi_{\bm{p}^{\prime}}}$, which is not enough to guarantee the interpretation of harmonics as vortex modes, as certain azimuth-angle-dependent phases are hidden in the electron and photon spinor and polarization vector. Moreover, in Eq.~(\ref{S_fi_NCS_pl}), the photon absorption number in the $\delta$-function coincides with the harmonic number in the complex phase, which is quite in contrast with the pulsed case.  

After assigning the zero-vortex mode for the final electron through $b_{\rho^\prime}(\bm{p}^\prime_\perp)$, the above $S$-matrix element would produce the wave function of the vortex photon in the following form:
\begin{eqnarray}
	\begin{aligned}
		&\bm{A}_{\bm{k}^{\prime},\Lambda^{\prime}}=\frac{\mathcal{N}^\prime_\text{{NCS}} }{(2\pi)^2}\sum_{n,\sigma_{\Lambda^{\prime}}} \delta(q_z + n k_z- q^{\prime}_z-{k}^{\prime}_z)\,\delta(q_0 + n k_0- q^{\prime}_0-{k}^{\prime}_0)\\
		&\times \sqrt{\frac{2\pi}{{k_{\perp}^{\prime}}}}\delta(k_{\perp}^{\prime}-\rho^{\prime}) (-1)^{m^\prime_n} e^{i(m^\prime_n-\sigma_{\Lambda^{\prime}})\varphi_{\bm{k}^{\prime}}} d_{\sigma_{\Lambda^{\prime}},\Lambda^{\prime}}^{1}(\theta_{\bm{k}^{\prime}})\,\bm{\chi}_{\sigma_{\Lambda^{\prime}}}{M}_{n}\,.
	\end{aligned}
	\label{evolved-momentum-pw}
\end{eqnarray}
Unlike the pulsed laser case, there is a one-to-one correspondence between the NCS harmonic mode and the vortex mode. Thus, $\omega^{\prime}$ and $\theta_{\bm{k}^{\prime}}$ would uniquely define the harmonic content of NCS in the monochromatic laser case. Therefore, in the monochromatic laser case, one just identifies the harmonics as vortex modes and the energy spectrum of the radiation is the same for the plane wave $\gamma$ and the  vortex $\gamma$ emissions. 

\begin{figure}[H]
	\begin{center}
		\includegraphics[width=0.95\linewidth]{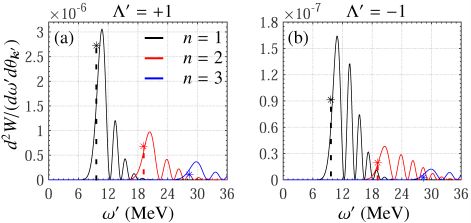}
		\begin{picture}(300,0)  
\end{picture}
		\caption{\small{Radiation rate for vortex photon with cone angle $\theta_{\bm{k}^{\prime}}\approx0.35$ mrad for the first 3 harmonics for photon helicities (a) $\Lambda^{\prime}= +1$ and (b) $\Lambda^{\prime}=- 1$. Asterisks correspond to the monochromatic laser case and solid lines correspond to the pulsed case. The monochromatic results have been rescaled by a constant factor to facilitate a better comparison with the pulsed result. }}  \label{fig_energy}
	\end{center}
\end{figure}

\begin{figure}[H]
	\begin{center}
		\includegraphics[width=0.95\linewidth]{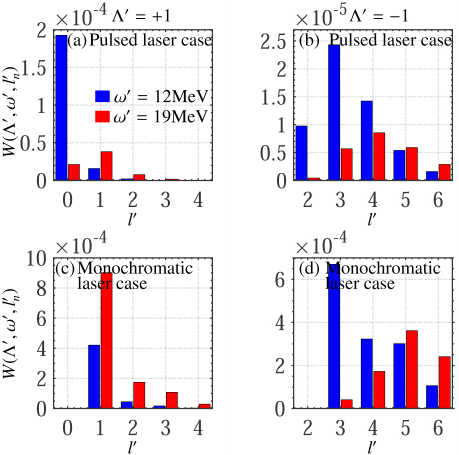}
		\begin{picture}(300,0)  		
		\end{picture}
		\caption{\small{The OAM distributions of incoherent vortex $\gamma$ photons with energies $\omega^\prime=12$ MeV and $\omega^\prime=19$ MeV for helicities [(a) and (c)] $\Lambda^{\prime}= +1$ and [(b) and (d)] $\Lambda^{\prime}=- 1$, respectively. The first and second lines correspond to pulsed and monochromatic laser cases respectively.} } \label{fig_oam}
	\end{center}
\end{figure}

\begin{figure}[H]
	\setlength{\abovecaptionskip}{-0.0cm}
	\includegraphics[width=1\linewidth]{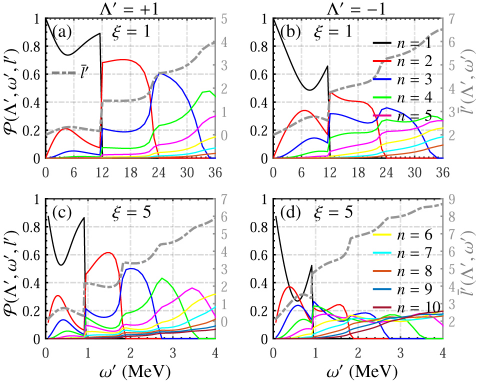}
	\begin{picture}(300,0)   
	\end{picture}
	\caption{OAM ratios $\mathcal{P}(\Lambda^\prime, \omega^\prime,l^\prime_n)$ over energy spectra for polarized vortex $\gamma$ photons generated in the monochromatic CP laser. }
	\label{fig_oam-ratio-mono}
\end{figure}

In the numerical calculations below, the electron energy is $\varepsilon=1$ GeV, and the laser photon energy is $\omega=1.55$ eV.
The difference between pulsed and monochromatic laser cases concerning vortex $\gamma$ radiation are shown in the energy spectrum and the OAM content in Figs. \ref{fig_energy} and \ref{fig_oam}.  As is shown in Fig.~\ref{fig_energy}, the spectra of the first harmonic spreads into the region where one expects the result of the second harmonic. In Fig.~\ref{fig_oam}, for the vortex $\gamma$ photon with energy $\omega^\prime=12$ MeV, one observes contributions from the fundamental harmonic [see Figs.~\ref{fig_oam} (a) and \ref{fig_oam} (b)] which is absent in the monochromatic laser case [see Figs.~\ref{fig_oam} (c) and \ref{fig_oam} (d)]. 

Similar OAM ratios of the polarized $\gamma$ photons, as defined in the main text for the pulsed laser case, can be introduced here for the monochromatic laser case: $\mathcal{P}(\Lambda^\prime, \omega^\prime,l^\prime_n)=\frac{W(\Lambda^\prime, \omega^\prime,l^\prime_n)}{\sum_{n} {W(\Lambda^\prime, \omega^\prime,l^\prime_n)}}$, where $W(\Lambda^\prime, \omega^\prime,l^\prime_n)$ is obtained from the radiation rate for the $n$th harmonic \cite{ivanov2004complete}.  Corresponding numerical results are presented in Fig.~\ref{fig_oam-ratio-mono}.  Not that the differences between results from the pulsed laser in the main text and the monochromatic laser case presented here:
(1) the lower harmonics cannot spread into higher harmonic regions, and the resulting average OAM has a staircase pattern, while in the pulsed case a smooth increment is observed; (2) the monochromatic laser case overestimates the OAM ratios due to the lack of spreading of the lower harmonics.

\bibliography{vblib}

\end{document}